\documentstyle[12pt]{article}
\begin{document}

\newcommand{\be}{\begin{equation}}
\newcommand{\ee}{\end{equation}}
\begin{flushright}
BRX-TH 415\\
\end{flushright}
\centerline{\large Accelerated Detectors and Temperature in (Anti) de Sitter
Spaces\\}

\vspace{0.3cm}
\centerline{ S. Deser and Orit Levin \\}
\vspace{0.3cm} 

\centerline{Physics Department, Brandeis University, Waltham, Massachusetts
02254, USA\\} 
\vspace{1cm}
{\large Abstract\\}

We show, in complete accord with the usual Rindler picture,
 that \linebreak
detectors with constant acceleration $a$
in de Sitter (dS) and Anti de Sitter (AdS) spaces 
with cosmological
constants $\Lambda$  measure temperatures \linebreak
$2\pi T=(\Lambda/3+a^{2})^{1/2}\equiv a_{5}$, the detector "5-acceleration"
in the embedding flat 5-space.
For dS, this recovers a known result;  
in AdS, where  $\Lambda$ is negative,
the temperature is well defined down to the critical value $a_{5}=0$,
again in accord with the underlying kinematics. 
The existence of
 a thermal spectrum is also demonstrated 
for a variety of candidate wave functions
in AdS backgrounds.\\

\vspace{0.5cm}

A  comoving detector in dS space 
 is well known \cite{birrell} to be in an effective
  thermal bath with temperature $2\pi T=(\Lambda/3)^{1/2}$ where $\Lambda$
is the (positive) cosmological constant, defined by
$R_{\mu\nu\alpha\beta}=\frac{1}{2}(g_{\mu\alpha}g_{\nu\beta}
-g_{\mu\beta}g_{\nu\alpha})\Lambda$.
 This result was recently extended  \cite{Narnhofer} to detectors with
 constant 4-acceleration $a$,  
whose temperature becomes  
  $2\pi T=(\Lambda/3+a^{2})^{1/2}$.
Our main purpose here is to investigate the extent to which  temperature is
applicable in AdS geometry,  despite its hyperbolicity problems,
but we will for unity provide parallel treatments of both dS and AdS spaces,
stressing in particular the usefulness of the 
embedding flat 5-spaces, in which the detector's 5-acceleration is precisely
$(\Lambda/3+a^{2})^{1/2}$, 
for understanding the desired observer motions. 
We will of course recover the above dS result; in AdS we find  the same
formula, as might be expected on formal continuation grounds.
However, the physics is quite different since there is a critical value 
 $a^{2}=-\Lambda/3$  of the acceleration below which the detector
 seems to measure imaginary temperature;
the resolution is of course that its motion becomes spacelike. That these 
simple kinematical considerations indeed lead to the correct physics
 will be verified  by the usual calculation of detector-quantum field 
interaction to exhibit a thermal distribution  . For AdS, we will consider
various boundary conditions that define states and show that they agree.\\

Our spaces  are represented as hyperplanes in flat embedding D=5 spaces
according to
\be \label{hds}
ds^{2}=\eta_{A \;B}dz^{A}dz^{B}\;\;\;\;\;\;\;\,\,
\eta_{A \; B}z^{A}z^{B}=\mp R^{2}
\ee
 $A,\;B =\; 0...4$,  $\eta_{A \;B}
={\rm diag} (1,-1,-1,-1,\mp 1)$ 
for dS/AdS respectively and $R^{2} \equiv 3|\Lambda|$. 
Now consider detectors moving according to
 $z^{2}=z^{3}=0$ \linebreak 
 $z^{4}=Z={\rm const}$; 
 their 5-space  motions are
 \be \label{flat1}
(z^{1})^{2}-(z^{0})^{2}=\pm(R^{2}-Z^{2})\equiv a_{5}^{-2}
\ee                          
which resemble the hyperbolic trajectories 
\be\label{flat}
x^{2}-t^{2}=a^{-2}
\ee
of  accelerated detectors 
in Minkowski space.
There, such  detectors measure
temperature $2\pi T=a$, so we would expect our observer (\ref{flat1})
to see one as well, with   
\be\label{temp}
2\pi T=\left(\pm (R^{2}- Z^{2})\right)^{-1/2}.
\ee
These dS/AdS sign correlations agree with the formal fact  
 that the spaces are related by a change of real to imaginary $R$ and
$z^{4}$, but of course this kinematic argument 
 has to be confirmed for both spaces, 
{\it e.g.}, by exhibiting suitable Planck distributions for quantum 
fields in these
backgrounds, and  we must also express and understand these motions in
the physical 4-space.\\

\vspace{3mm}

Consider first dS, expressed in  the standard coordinates covering
the whole manifold
\be
\label{dsmetric}
ds^{2}=dt^{2}-R^{2}\cosh^{2}t/R \left[
d\chi^{2}+\sin^{2}\chi
\left( d\theta^{2}+\sin^{2}\theta d\phi^{2}\right) \right],
\ee
which are related to the $z^{A}$ of SO(1,4) according to
\begin{eqnarray}\label{coor}
z^{0}\!\!&=&\!\!\!R\sinh t/R
\;\;\; , \;\;\;\;
z^{4}=R\cos\chi\cosh t/R \;\;\; , \;\;\; z^{1}=R\sin\chi\cos\theta\cosh t/R
\nonumber \\
z^{2}\!\!&=&\!\!R\sin\chi\sin\theta\cos\phi\cosh t/R\;\;\; ,\;\;\;
z^{3}=R\sin\chi\sin\theta\sin\phi \cosh t/R. 
\end{eqnarray}
For a radially moving detector, we may always (by the symmetry) choose
\linebreak $\theta=0\, ,\,\phi={\rm const}$ or
$z^{2}=z^{3}=0$.
We further set  $z^{4}=Z$, a constant or
since  $z^{1}$ and $z^{4}$ are on the same footing, another set of 
trajectories with the same acceleration is given by
 $z^{1}=Z={\rm const}$. The two corresponding sets of (timelike)
trajectories are
illustrated in Fig 1.
Because $|\cos\chi|\leq 1$, this means that $Z^{2}\leq R^{2}\cosh^{2}t/R$,
hence trajectories with  $|Z|<R$ are timelike, $|Z|>R$ spacelike, and 
$|Z|=R$  null; we consider only $|Z|<R$
Next, let us evaluate the magnitude  
 $a^{2}=-g_{\beta\gamma}a^{\beta}a^{\gamma}$ of the 4-acceleration
$D^{2}x^{\mu}/Ds^{2}$.
A straightforward calculation yields the
 constant value
\be \label{ads}
a^{2}R^{2}=Z^{2}/(R^{2}-Z^{2}).
\ee 
Inserting $Z^{2}=a^{2}R^{4}(1+R^{2}a^{2})^{-1}$ into (\ref{temp}),
or  using the relation $a_{5}^{2}=R^{-2}+a^{2}$ between the 5- and 
4-accelerations (valid for every   
 radial trajectory),
yields the result of \cite{Narnhofer}
\be \label{tds}
2\pi T=
\left(R^{-2}+a^{2}\right)^{1/2}.
\ee 
The temperature found in  (\ref{tds}) is also independent
of the particular coordinates chosen.
Thus it is the effective 
"Unruh acceleration" $a_{5}$ of the detector in the embedding
flat 5-space that determines the temperature.\\
The fundamental justification 
for the applicability of  temperature  is of
course to be found in the form taken by quantum fields on the dS 
background.
The Wightman function for conformally coupled massless
scalar field has been shown  \cite{Tagirov} (see also \cite{birrell})
 to be identical to that
in  flat space, namely proportional to the square of the invariant 
5-distance between the two points on the detector trajectory there.
Hence an "Unruh effect" calculation has to give the same result as in flat
space but now in terms of the 5-acceleration relevant in the embedding
space. Thus the QFT justification of our result is immediate from this
point of view. Equally convincing, of course, is the derivation 
of \cite{Narnhofer} based
on the KMS interpretation of 
quantized fields \cite{Bros} in dS.\\

We turn now to AdS, 
parametrized according to
\be 
ds^{2}=R^{2}(\cos \rho)^{-2} \left[ d\tau^{2}-
d\rho^{2}
-\sin^{2}\rho(d\theta^{2}+\sin^{2}\theta d\phi^{2})\right]
\ee   
using 
\begin{eqnarray}\label{coor2}
 z^{0}\!\!\!\!&=&\!\!\!\!R\sin \tau/\cos \rho \;\;\;\;\;
  z^{4}=R\cos \tau/\cos \rho \;\;\;\;\;
 z^{1}=R \tan \rho\cos \theta \nonumber\\
z^{2}\!\!\!\!&=&\!\!\!\!R\tan\rho\sin\theta\cos\phi\;\;\;\;\;
z^{3}=R\tan\rho\sin\theta\sin\phi ;
\end{eqnarray}
the ranges $0\leq\rho<\pi/2, \, 0\leq\theta\leq \pi,\, 0\leq\phi<2\pi$
cover the whole manifold. 
We consider, as before, trajectories with 
$\theta\!=\!0,\, \phi\!=\!{\rm const}$, 
 so that $z^{2}=z^{3}=0$, and the effective 
interval reduces to
\be
ds^{2}=R^{2}(\cos \rho)^{-2}(d\tau^{2}-d\rho^2).
\ee    
A motion with $z^{1}$ or $\rho=\rm{const}$
  experiences the constant
4-acceleration 
\be
a^{2}R^{2}=\sin^{2}\rho
\ee
which is maximized  
at  spatial infinity, $\rho=\pi/2$, and vanishes  at $\rho=0$. 
\linebreak However, lines of constant $\rho$ are now realized 
 as the circles \linebreak
$(z^{0})^{2}+(z^{4})^{2}={\rm const}$ in the embedding 5-space;
these no longer correspond to the 
 hyperbolic trajectories a la Minkowski,
and do not immediately suggest any thermal properties; we shall see below that
they correspond to $T=0$.
In contrast, we focus on the timelike lines 
$z^{4}=Z$ with  $|Z|>R$  
($|Z|<R$ are spacelike) that  do follow 
 hyperbolic trajectories,
\be
(z^{1})^{2}-(z^{0})^{2}=Z^{2}-R^{2}=a_{5}^{-2}
\ee
 with constant 4-acceleration
\be\label{aa}
a^{2}R^{2}=Z^{2}\left((Z^{2}-R^{2})\right)^{-1}.
\ee
Here $a$ ranges 
 from $R^{-1}$ for $|Z|=\infty$
to infinity as $|Z|$ decreases from $\infty$ to $R$.
Figure 2 exhibits the constant $z^{4}$ lines 
 for AdS.
The line $Z^{2}=\infty$ is  $\rho=\pi/2$, while 
the null lines $\tau=\pm\rho+n\pi$ 
are the $Z^{2}=R^{2}$   horizons: no detector  
can cross them.
[We could have chosen an equivalent set of observers
by setting $z^{0}$ rather than $z^{4}$ constant.
These are  translated by $\pi/2$ in the $\tau$ direction with respect to 
the $z^{4}=Z$ choice; the  horizons 
lines are likewise moved by $\pi/2$.]
We therefore expect, from the Minkowski-Rindler analogy in the flat 
embedding space description
of detector motion, 
   the AdS  
detectors on the $z^{4}=Z$ paths to measure the temperature  
\be
2\pi T=\left(-R^{2}+ Z^{2}\right)^{-1/2}=
\left(-R^{-2}+a^{2}\right)^{+1/2}.
\ee  
As in dS, the relation $a_{5}^{2}=-R^{-2}+a^{2}$ actually holds for every
radial trajectory, but of course $T$ is only defined for constant
accelerations.
Since physical detectors cannot travel on 
 the spacelike 4-orbits with $Z^{2}<R^{2}$,
we need not worry about imaginary $T$. 
 
Having motivated the existence and the form of the AdS temperature
formula, we must now attempt to derive it.
We use the Unruh approach -the excitation of a detector by a 
quantum field in the AdS background- thereby
  determining the temperature by the (lowest order) transition rate 
obtained from the Wightman function.
Unlike the dS case, however there is an
{\it a priori} problem: the lack of global hyperbolicity
of AdS.
We have nothing to add to this question, and will
  simply take over the results of
\cite{Avis} for defining a massless scalar field 
in a "causal" way, using the equivalent
of box quantization.
There is an ambiguity in defining the wave function $\Phi$, depending on the
 assumed
boundary conditions,  in terms of the wave function $\Phi_{E}$
 of the static Einstein universe (into which AdS may be conformally mapped).
The choices discussed in \cite{Avis} 
are given respectively by $\Phi_{A} =(\cos\rho/R)\Phi_{E}$ 
("transparent" boundary conditions), and two "reflective" conditions 
characterized  by
choosing $\Phi_{B},\, \Phi_{C}$ such that they respectively 
obey the "wall" conditions 
$\partial\Phi_{E}
/\partial\rho \rightarrow 0$ 
or $\Phi_{E}
\rightarrow 0$ at $\rho \rightarrow \pi/2$.
For each of these 3 choices, we can compute the Wightman function to obtain
\be
 8\pi^{2}R^{2}W_{n}(x,x')/_{\theta=\theta '\,\phi=\phi '}=
X[1+\varepsilon_{n}(1+2X)^{-1}]
\ee
\[ X=\frac{\cos\rho\cos\rho'}{
[\cos(\tau-\tau'-i\epsilon)-\cos(\rho-\rho')} 
\;\;\;\;\;\;\;\;\; (\varepsilon_{A}, 
    \varepsilon_{B}, \varepsilon_{C})=(0,-1,1) \]
for cases A,B,C respectively.     
For $\rho={\rm const}$ trajectories, $\tau=s \cos\rho/R$ 
where $s$ is  proper time, so  $W$
depends only on proper time differences.
The  corresponding transition rate  is then  of the form
\be \label{Gamma}
\Gamma_{n}=A\int^{\infty}_{-\infty}d(s-s')\exp[-i\Delta E(s-s')]\,W_{n}(s-s');
\ee
 $A$ is a constant depending on the detector-scalar
field interaction and $\Delta E$ the energy difference between detector
energy levels.
We can  immediately check that the integral  vanishes for $\rho={\rm const}$ 
trajectories 
  (all the pole in $W_{n}$ are in the upper half-plane)
so that temperature vanishes, 
in agreement with the non-hyperbolic nature 
of these 
paths  noted above.
For the $z^{4}=Z$ trajectories we can solve for the proper time
and express everything in terms of it and the acceleration to find that
\be
X=a_{5}^{2}R^{2}[1-\cosh a_{5}(s-s')-i\epsilon]^{-1}.
\ee
Hence $X$ is a function of $(s-s')$ 
only and vanishes as $(s-s')\rightarrow \infty$,
 so we may use the adiabatic transtion rate  (\ref{Gamma}).
The contribution of $X$ to $\Gamma$ is
\be
\Gamma_{A}=\frac{A\Delta E}{2\pi}[\exp(
2\pi\Delta E /a_{5})-1 ]^{-1},
\ee
the same expression as for the constant acceleration detector in Minkowski
space \cite{birrell}, so that our original expectation is fulfilled .
The other two cases differ from A by extra terms, which after some arithmetic
are found to be 
\be
\Gamma_{B,C}=\Gamma_{A}\pm\frac{A}{2\pi}\frac{a_{5}\sin[b\Delta E
/a_{5}]}{\sinh b}
[ \exp(
2\pi\Delta E / a_{5})-1]^{-1 } 
\ee
where $b\equiv 1+2a_{5}^{2}R^{2}\geq 1$. For fixed $a_{5}$, 
the limit as $R\rightarrow \infty$
(flat space) of these extra terms reassuringly vanishes.
More generally, we argue that the dominant term here is 
still the Planck denominator,
so that the notion of temperature  persists, as does its magnitude also 
for the B and C boundary conditions.\\
 
In summary, we have established a Rindler-like kinematical basis for
associating temperature with constant accelerated motions in AdS (as in dS)
 geometries, their
effective accelerations in the embedding flat 5-space being the
relevant parameter.
In AdS, this picture also accounted for the existence of a threshhold value 
of the 
4-acceleration above which temperature is well-defined. These considerations
were confirmed by explicit calculations of the correlators of a quantum field 
in AdS, whose result was independent of the  different boundary conditions 
permitted 
by the AdS causality ambiguities, and led to the standard Unruh  Planckian
temperature distribution.\\

SD  is happy to thank the authors of \cite{Narnhofer} for 
stimulating  conversations. This work was supported by NSF grant
PHY-9315811, OL by the Fishbach Foundation.

\newpage
{\bf Fig 1} Penrose diagram of dS space. Solid lines represent
 $z^{4}=\pm|Z|$, broken lines the two possible sets with 
$z^{1}=Z$.  
Dotted lines give
the null horizons of the trajectory sets to which they are the asymptotes. \\

\vspace{1cm}

{\bf Fig 2} AdS space. Solid lines represent 
 different sets of  $z^{4}=Z$ trajectories:  $ Z$ negative
 for $-3\pi/2<\tau<-\pi/2 $, $Z$ positive
for $-\pi/2<\tau<\pi/2$ , etc, alternating in each $\pi$-interval.
The dotted lines are the corresponding null horizons.

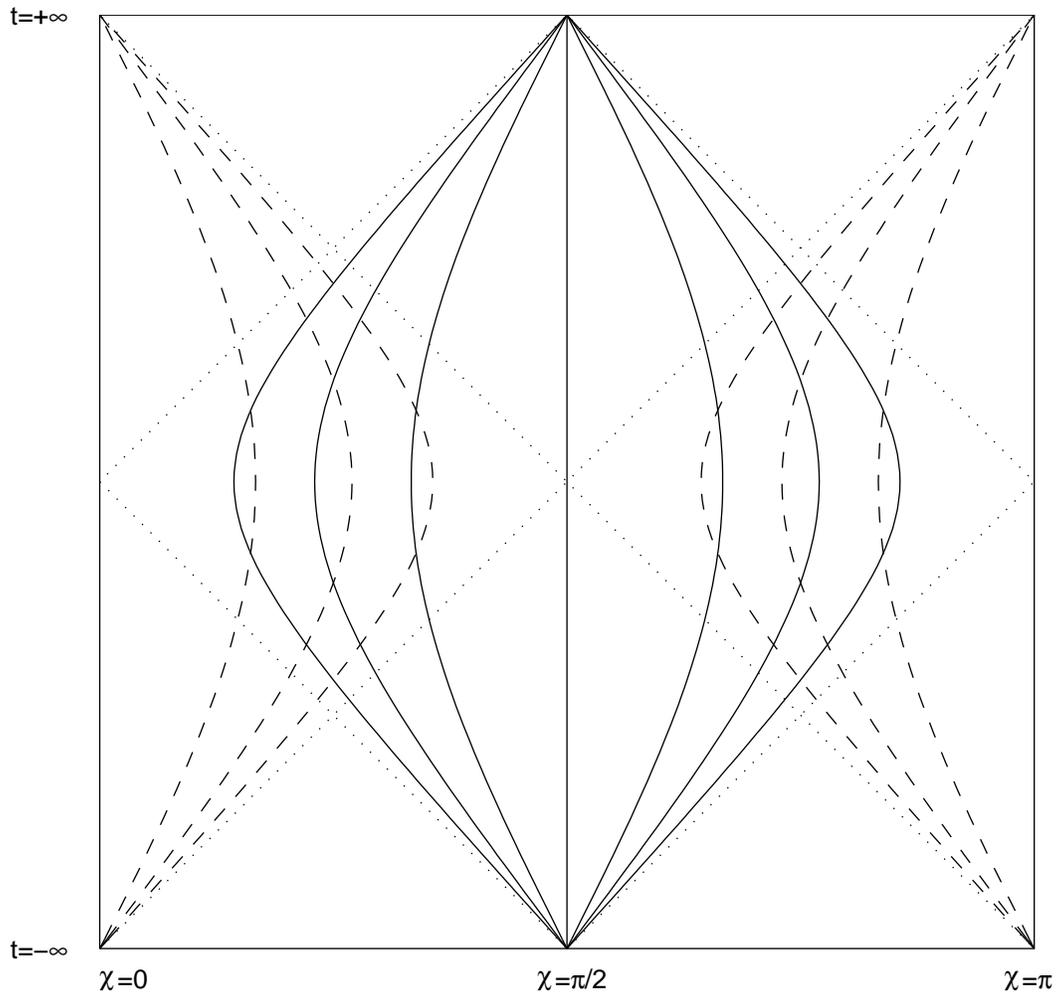
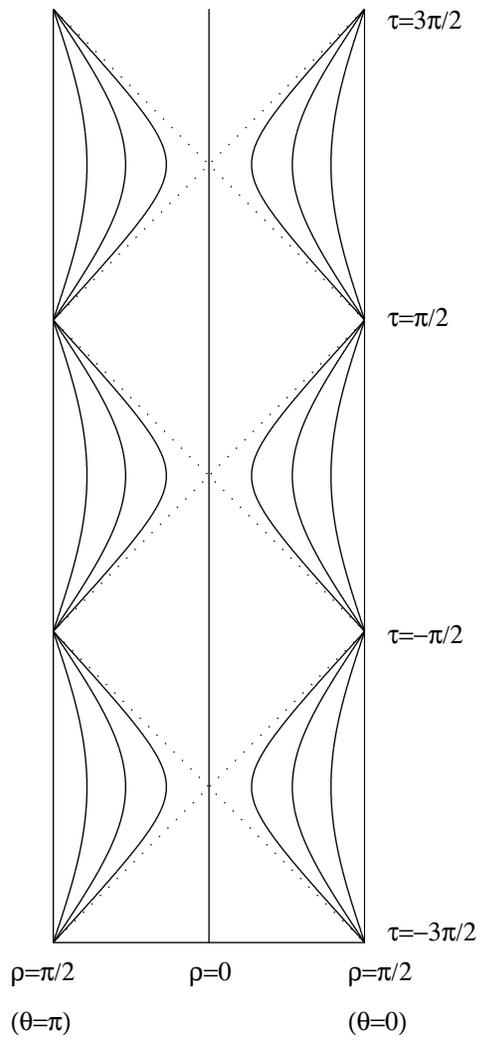
\end{document}